\documentstyle[seceq,preprint,epsfig]{jpsj} 
\renewcommand\figureheight[1]{\vspace{24pt}\mbox{\rule{0cm}{#1}}}

\newcommand{\vv}[1]{\mbox{\boldmath{$#1$}}}
\newcommand{\vs}[1]{\mbox{\footnotesize \boldmath{$#1$}}}

\def\neq{ = \hskip -1.0em / \hskip 0.6em}

\def\runtitle{Peierls Distortion in 2D Tight-Binding Model} 
\def\runauthor{Y. {\sc Ono} 
and T. {\sc Hamano}} 

\title {Peierls Distortion in Two-Dimensional Tight-Binding Model} 
\author {Yoshiyuki {\sc Ono}\footnote{E-mail : ono@ph.sci.toho-u.ac.jp} 
and Tetsuya {\sc Hamano}\footnote{E-mail : hamano@ph.sci.toho-u.ac.jp}} 
\inst { Department of Physics, Toho University, Miyama 2-2-1, Funabashi, 
Chiba 274-8510 } 
\recdate {January 26, 2000} 
\abst {The Peierls distortions in a two-dimensional electron-lattice 
system described by a Su-Schrieffer-Heeger type model extended to 
two-dimensions are numerically studied for a square lattice. The 
electronic band is just half-filled and the nesting vector is 
($\pi/a$, $\pi/a$) with $a$ the lattice constant. In contrast 
to the previous understanding on the Peierls transition in two dimensions, 
the distortions which are determined so as to minimize the total energy 
of the system involve not only the Fourier component with the nesting 
wave vector but also many other components with wave vectors parallel 
to the nesting vector. It is found that such unusual distortions 
contribute to the formation of gap in the electronic energy spectrum 
by indirectly (in the sense of second order perturbation) connecting two 
states having wave vectors differing by the nesting vector from each other. 
Analyses for different system sizes and for different electron-lattice 
coupling constants indicate that the existence of such distortions is not 
a numerical artifact. It is shown that the gap of the electronic energy 
spectrum is finite everywhere over the Fermi surface.
} 

\kword{Peierls transition, two-dimensional electron-lattice systems, 
nesting, Peierls gap, Peierls distortion, second order perturbation}

\begin{document} 
\sloppy 

%\twocolumn

\maketitle 

%\newpage
%
%\begin{multicols}{2}
%%%%%%%%%%%%%%%%%%%%%%%%%%%%%%%%%%%%%%%%%%%%%%%%%%%%%%%%%%%%%%%%%%%%%%%%%%%%
%%%%%%%%%%%%%%%%%%%%%%%%%%%%%%%%%%%%%%%%%%%%%%%%%%%%%%%%%%%%%%%%%%%%%%%%%%%%

%\begin{large}

\section{Introduction} 

The Peierls transition is caused by the freezing of a lattice distortion 
mode which can connect degenerate electronic states at the Fermi 
level.\cite{peierls} The presence of such a distortion induces an energy 
gap at the Fermi level of the electronic spectrum. This gap which is called 
Peierls gap lowers the electronic energy. In some cases, this reduction of 
energy overcomes the increase of the lattice energy due to the frozen mode. 
In one-dimensional systems, particularly, the lowering of the electronic 
energy is proportional to the square times logarithm of the frozen mode 
amplitude and therefore can overcome the lattice energy increase which is 
proportional to the square of the amplitude. Because of the competition 
between the decrease of the electronic energy and the increase of the 
lattice energy, there exists a value of the frozen mode amplitude minimizing 
the total energy of the electron-lattice system. 

In dimensions higher than one, the situation is not so simple. This is 
because the number of states at the Fermi level is only two in the case 
of a one-dimensional system, whereas those in two- or three-dimensional 
systems form equi-energy line or surface, respectively. In general, a 
single mode of the lattice distortion can connect only two points in 
the Fermi surface (or line). In this situation, the gain in the electronic 
energy is too small to overcome the increase of the lattice energy. 
However, in some special situation where a single lattice distortion 
mode can connect many states at the Fermi level, the electronic 
energy is lowered substantially and the Peierls transition becomes 
possible. This situation is known as ^^ ^^ nesting''. The simplest 
case can be seen in the two-dimensional square lattice tight-binding 
model with a half-filled electronic band. In this case, the Fermi 
line is a square within the first Brillouin zone combining four points, 
$(\pi/a,0)$, $(0,\pi/a)$, $(-\pi/a,0)$ and $(0,-\pi/a)$, with $a$ the 
lattice constant. The nesting vector of this system is 
$\vv{Q}=(\pi/a,\pi/a)$; $\overline{\vv{Q}}=(\pi/a,-\pi/a)$ is also a 
nesting vector, but it is equivalent to $\vv{Q}$. 

\begin{figure}[htb]
\figureheight{5.2cm}
%\vspace*{-5cm}
%\epsfysize=5.5cm
\hspace*{38mm}
\epsfig{file=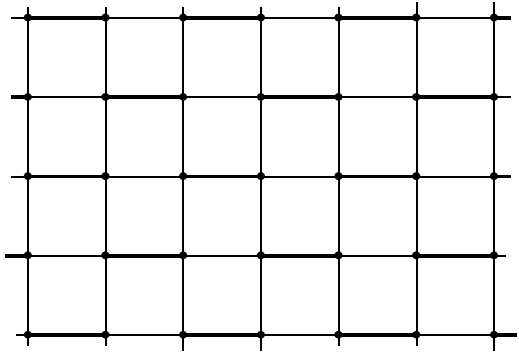,width=260pt,height=174pt}
%\centerline{\epsfbox{hirsch1.ps}}
%\vspace*{5cm}
\caption{The asymmetric dimerization pattern. The bonds with thick horizontal 
lines represent shorter bonds, and the other horizontal bonds are longer 
ones. There is no dimerization in the perpendicular direction in this 
situation. }
\label{fig1}
\end{figure}

The Peierls distortion was extensively studied in the late eighties 
to the early nineties in connection to the high $T_{\rm c}$ 
superconductors.\cite{machida87,hirsch1,scal89,mazumdar} 
A typical argument was given, e.g., by Tang and Hirsch\cite{hirsch1} 
who used a tight-binding model where the transfer integral is modified 
by the lattice displacement; its one-dimensional version is known as 
Su-Schrieffer-Heeger's (SSH) model.\cite{SSH} According to the analysis 
given in ref.~\citen{hirsch1}, the lattice distortion minimizing the 
total energy has the wave vector $\vv{Q}$ and a polarization parallel 
to one of the main crystal axes as shown in Fig.~\ref{fig1}. 

As will be discussed in this paper, the straight forward numerical 
study of the ground state of the tight-binding square lattice 
electron-lattice system described by the 2D version of the SSH model 
indicates that the distortion pattern shown in Fig.~\ref{fig1} does not give 
the lowest energy state. Fourier analysis of the distortion pattern 
yielding the lowest energy state shows the freezing of many other 
modes with wave vectors parallel to $\vv{Q}$ in addition to that of 
the main mode with $\vv{Q}$. We discuss in this paper why this is 
the case, along with the results of numerical calculations. 

In the next section, the model and the method of determining the lowest 
energy state are described. In \S~3, the reason why many modes with wave 
numbers different from $\vv{Q}$ can contribute to lower the total energy 
will be discussed.  Based on the discussion in \S~3, we will propose in 
\S~4 a new method to find the lowest energy state, which can reduce a single 
2D problem essentially to many 1D problems. The last section is devoted 
to summary and discussion.

%%%%%%%%%%%%%%%%%%%%%%%%%%%%%%%%%%%%%%%%%%%%%%%%%%%%%%%%%%%%%%%%%%%%%%%%%  
%%%%%%%%%%%%%%%%%%%%%%%%%%%%%%%%%%%%%%%%%%%%%%%%%%%%%%%%%%%%%%%%%%%%%%%%%
\section{Model and Basic Formulation}

The model Hamiltonian treated in this paper is the following 
SSH-type  Hamiltonian extended to two dimensions, 
\begin{eqnarray} 
H =  &- &\ \sum_{i,j,s}[ (t_0 -\alpha x_{i,j})(c_{i+1,j,s}^{\dagger} 
c_{i,j,s} + {\rm h.c.}) \nonumber \\ 
& & +(t_0 -\alpha y_{i,j})(c_{i,j+1,s}^{\dagger} 
c_{i,j,s} + {\rm h.c.})] \nonumber \\
&+ & \frac{K}{2} \sum_{i,j} (x_{i,j}^2+y_{i,j}^ 2), 
\label{eq2.1}
\end{eqnarray}
where $(i,j)$ represents a lattice point in a square lattice 
with a lattice constant $a$, 
and $x_{i,j} = u_x(i+1,j)-u_x(i,j)$, $y_{i,j} = u_y(i,j+1)-u_y(i,j)$ with 
$\vv{u}(i,j)$ the displacement vector of an ion-unit at the site $(i,j)$, 
$t_0$ the transfer integral of the equidistant (undeformed) lattice, $\alpha$ 
the electron-lattice coupling constant, $c_{i,j,s}^{\dagger}$ and $c_{i,j,s}$ 
the creation and annihilation operators of an electron at the site $(i,j)$ and 
with spin $s$. The last term on the right hand side of eq.~(\ref{eq2.1}) 
describes the lattice harmonic potential energy with $K$ the force constant. 
The lattice kinetic energy is omitted since we do not discuss the dynamical 
problem in this paper. 

If we assume the Peierls distortion having the nesting vector 
$\vv{Q} = (Q_x,\ Q_y) = (\pi/a,\ \pi/a)$ as done by Tang and 
Hirsch,\cite{hirsch1} the bond variables $x_{i,j}$ and $y_{i,j}$ 
can be expressed in the following form,
\begin{eqnarray}
x_{i,j} & = & x_0{\rm e}^{{\rm i}(Q_xi+Q_yj)a} = (-1)^{i+j}x_0 ,  
\label{eq2.2} \\
y_{i,j} & = & y_0{\rm e}^{{\rm i}(Q_xi+Q_yj)a} = (-1)^{i+j}y_0,   
\label{eq2.3}
\end{eqnarray}
where $x_0$ and $y_0$ are the amplitudes of the distortion to be determined 
to minimize the total energy of the system. In this situation the electronic 
part of the above Hamiltonian can be easily diagonalized by introducing the 
Fourier expansion of the electronic field operators as follows, 
\begin{equation}
c_{i,j} = {1 \over N}\sum_{\vs{k}} c_{\vs{k}}{\rm e}^{{\rm i}(k_xi+k_yj)a}, 
\label{eq2.4}
\end{equation}
where $N^2$ is the total number of lattice points and the components of the 
wave vector $\vv{k} = (k_x,k_y)$ are given by integer times $2\pi/Na$, 
respectively, on the assumption of the periodic boundary conditions. 
The diagonalization process is straightforward, and the ground state 
energy of the system is expressed in the form, 
\begin{equation}
E_{\rm tot}^{\rm GS} = -2{\sum_{\vs{k}}}' E_{\vs{k}} + 
N^2{K \over 2}(x_0^2+y_0^2) , \label{eq2.5}
\end{equation}
where the sum over the wave vector $\vv{k}$ is restricted to the region 
satisfying the following two conditions,
\begin{eqnarray}
-{\pi \over a} < & k_x + k_y & \le {\pi \over a} \label{eq2.6} \\
-{\pi \over a} < & k_x - k_y & \le {\pi \over a} \label{eq2.7}
\end{eqnarray}
and $E_{\vs{k}}$ is given by
\begin{equation}
E_{\vs{k}} = \sqrt{ \varepsilon_{\vs{k}}^2+\Delta_{\vs{k}}^2 } \label{eq2.8}
\end{equation}
with $\varepsilon_{\vs{k}} = -2t_0(\cos k_xa +\cos k_ya)$ and 
$\Delta_{\vs{k}}=2\alpha (x_0\sin k_xa + y_0\sin k_ya)$. The factor 2 
in front of the $\vv{k}$-sum in eq.~(\ref{eq2.5}) is due to the spin 
degeneracy. The boundary 
determined by the conditions eqs.~(\ref{eq2.6}) and (\ref{eq2.7}) is 
nothing but the Fermi surface (or line since the present system is two 
dimensional) and on this boundary $\varepsilon_{\vs{k}}$ vanishes. 
Across this boundary the electronic energy band is divided into two 
pieces, the dispersions of the lower and upper bands being expressed by 
$-E_{\vs{k}}$ and $E_{\vs{k}}$, respectively. This means that the 
energy gap at the Fermi surface is given by $2|\Delta_{\vs{k}}|$. 

According to Tang and Hirsch,\cite{hirsch1} the dimerization pattern shown 
in Fig.~\ref{fig1} which is realized by setting $x_0\neq 0$ and $y_0=0$ 
or vice versa and therefore highly asymmetric in $x$ and $y$ directions can 
have lower energy than the symmetric dimerization realized when 
$x_0=y_0 \neq 0$ (or equivalently $x_0=-y_0$). It is not difficult to 
understand the reason if we see the expression of $\Delta_{\vs{k}}$ for each 
case; in the asymmetric case, 
$\Delta_{\vs{k}}^{\rm asym} = 2\alpha x_0 \sin k_xa$, and in the symmetric 
case, $\Delta_{\vs{k}}^{\rm sym} = 4\alpha x_0 \sin [{1 \over 2}(k_x+k_y)a] 
\cos[{1 \over 2}(k_x-k_y)a]$. The gap on the Fermi surface does not vanish 
except for the points $(0,\pi/a)$ and $(\pi/a,0)$ in the case of the asymmetric 
dimerization.  On the other hand the gap vanishes on the line $k_x-k_y=\pi/a$ 
in the case of symmetric dimerization. We have confirmed numerically that 
if we fix the ratio $r=y_0/x_0$ ($0 \le r \le 1$) and minimize the total 
energy with respect to $x_0$, then the minimum value of the energy is a 
monotonically increasing function of $r$ and becomes the smallest at $r=0$. 

It is clear that the dimerization pattern shown in Fig.~\ref{fig1} can 
yield the lowest energy state as far as we consider only the distortion 
with the basic nesting vector $\vv{Q}$. However, we should note that 
even in this asymmetric dimerization the gap vanishes at some special 
points on the Fermi surface. In general the Peierls gap is proportional 
to the matrix element of the electron-lattice coupling term in $H$ between 
two electronic states with wave vectors $\vv{k}$ and $\vv{k}\pm \vv{Q}$. 
In the case of the SSH-type Hamiltonian as used in this paper, this matrix 
element is given by a linear combination of $\sin k_xa$ and $\sin k_ya$. 
It vanishes at $(\pi/a,0)$ and $(0,\pi/a)$ irrespectively of the lattice 
dimerization pattern. This means that the degeneracy between $(\pi/a,0)$ 
and $(0,-\pi/a)$ [or between $(0,\pi/a)$ and $(-\pi/a,0)$] cannot be removed 
within the first order perturbation due to the lattice distortion with 
$\vv{Q}$. We shall come back to this problem in \S~4.

%%%%%%%%%%%%%%%%%%%%%%%%%%%%%%%%%%%%%%%%%%%%%%%%%%%%%%%%%%%%%%%%%%%%%%%%%%%
%%%%%%%%%%%%%%%%%%%%%%%%%%%%%%%%%%%%%%%%%%%%%%%%%%%%%%%%%%%%%%%%%%%%%%%%%%%
\section{Numerical Study} %sec3

In order to check whether the asymmetric dimerization pattern shown in 
Fig.~\ref{fig1} can yield the lowest energy state or whether there exist 
any different dimerization patterns giving still lower energy, we have 
studied numerically the lowest energy state of the Hamiltonian 
eq.~(\ref{eq2.1}). Once we know the local values of $x_{i,j}$'s and 
$y_{i,j}$'s, the electronic wave functions \{$\phi_\nu(i,j)$\} are 
calculated along with corresponding eigenenergies \{$\varepsilon_\nu$\} 
from the following Schr\"odinger equation,
\begin{eqnarray}
\varepsilon_\nu \phi_\nu(i,j) & = & -(t_0-\alpha x_{i,j})\phi_\nu(i+1,j)
\nonumber \\
& & -(t_0-\alpha x_{i-1,j})\phi_\nu(i-1,j)  \nonumber \\
& & -(t_0-\alpha y_{i,j}) \phi_\nu(i,j+1) \nonumber \\
& & -(t_0-\alpha y_{i,j-1})\phi_\nu(i,j-1)] . \label{eq3.1}
\end{eqnarray}
Since the number of electrons is fixed, it is straightforward to obtain the 
electronic ground state energy for the given configurations of $x_{i,j}$'s 
and $y_{i,j}$'s. The bond length variables $x_{i,j}$'s and $y_{i,j}$'s are 
also involved in the lattice potential energy, and they are determined 
so as to minimize the total energy of the system. This condition yields 
the following self-consistent equations for these variables similarly as 
in one-dimensional cases,\cite{stafstrom,TO86} 
\begin{eqnarray}
x_{i,j} & = & -{{2\alpha} \over K}{\sum_\nu}' \phi_\nu(i+1,j)\phi_\nu(i,j) 
\nonumber \\
& & +{{2\alpha} \over {NK}}\sum_{i'}{\sum_\nu}' 
\phi_\nu(i'+1,j)\phi_\nu(i',j) , \label{eq3.2} \\
y_{i,j} & = & -{{2\alpha} \over K}{\sum_\nu}' \phi_\nu(i,j+1)\phi_\nu(i,j) 
\nonumber \\
& & +{{2\alpha} \over {NK}}\sum_{j'}{\sum_\nu}' 
\phi_\nu(i,j'+1)\phi_\nu(i,j'), \label{eq3.3}
\end{eqnarray}
where the summation over the one particle states $\nu$ is restricted to 
the occupied ones in the electronic ground state; note that the spin 
degeneracy factor should be included. The second terms on the right 
hand sides are due to the periodic boundary conditions, which require 
$\displaystyle{\sum_{i=1}^N x_{i,j} = 0}$ for arbitrary $j$ and 
$\displaystyle{\sum_{j=1}^N y_{i,j} = 0}$ for arbitrary $i$. 

In most of the practical calculations we use typically the following values 
of parameters; $t_0 = 2.5$eV, $K = 21.0$eV/\AA$^2$ and $\alpha=4.0$eV/\AA. 
These values are near to those for polyacetylene. 
As is well known what is important in this type of electron-lattice 
systems is the dimensionless coupling constant defined by $\lambda = 
\alpha^2/Kt_0$. Aforementioned values of parameters give 
$\lambda \simeq 0.30$. When we want to study the coupling constant 
dependence of various properties, only the value of $\alpha$ is changed 
for simplicity.

The set of self-consistent equations (\ref{eq3.1}) to (\ref{eq3.3}) can 
be solved numerically by iteration. First we give initial values for 
$x_{i,j}$'s and $y_{i,j}$'s, and then calculate $\phi_\nu(i,j)$'s by a 
matrix diagonalization subroutine, the results of which are substituted 
into the right hand sides of eqs.~(\ref{eq3.2}) and (\ref{eq3.3}). The 
resulting values of $x_{i,j}$'s and $y_{i,j}$'s are compared with the 
previous values. If the difference is not small, we proceed by replacing 
the initial values of $x_{i,j}$'s and $y_{i,j}$'s by new ones. This procedure 
is continued until the difference becomes negligibly small. 

We try three different initial configurations. First one is the asymmetrically 
dimerized pattern as shown in Fig.~\ref{fig1}. If we start the iteration with 
the uniform dimerization as expressed by eqs.~(\ref{eq2.2}) and (\ref{eq2.3}) 
by setting $y_0=0$ and giving an appropriate value of the order of 
$10^{-2}a$ for $x_0$, we end up with 
the same pattern with a value of $x_0$ which minimizes the total energy 
of the system within that pattern. The second choice is the symmetrically 
dimerized pattern as given by eqs~(\ref{eq2.2}) and (\ref{eq2.3}) with 
$x_0=y_0 \ne 0$. The third  one is a random distortion; we give random 
values for the lattice displacement vectors \{$\vv{u}(i,j)$\} with a 
relatively small amplitude of the order of $10^{-2} \times a$; many samples 
are studied in this treatment. The last two choices yield essentially the 
same result. 
The resulting distortion pattern is not described by eqs.~(\ref{eq2.2}) and 
(\ref{eq2.3}). An example of Fourier analysis of this pattern is shown in 
Fig.~2, where $X(q_x,q_y)$ and $Y(q_x,q_y)$ are defined by
\begin{eqnarray}
X(q_x,q_y) & = & {1 \over N^2}\sum_{i,j} x_{i,j}\exp[-{\rm i}(q_xi+q_yj)a] , 
\label{eq3.4} \\
Y(q_x,q_y) & = & {1 \over N^2}\sum_{i,j} y_{i,j}\exp[-{\rm i}(q_xi+q_yj)a] .
\label{eq3.5}
\end{eqnarray}
We have found that all the Fourier components with $q_x \ne q_y$ 
vanish.~\cite{footnote1}

\begin{figure}[htb]
%\vspace*{-5cm}
%\epsfysize=6.0cm
\figureheight{6cm}
\hspace*{30mm}
\epsfig{file=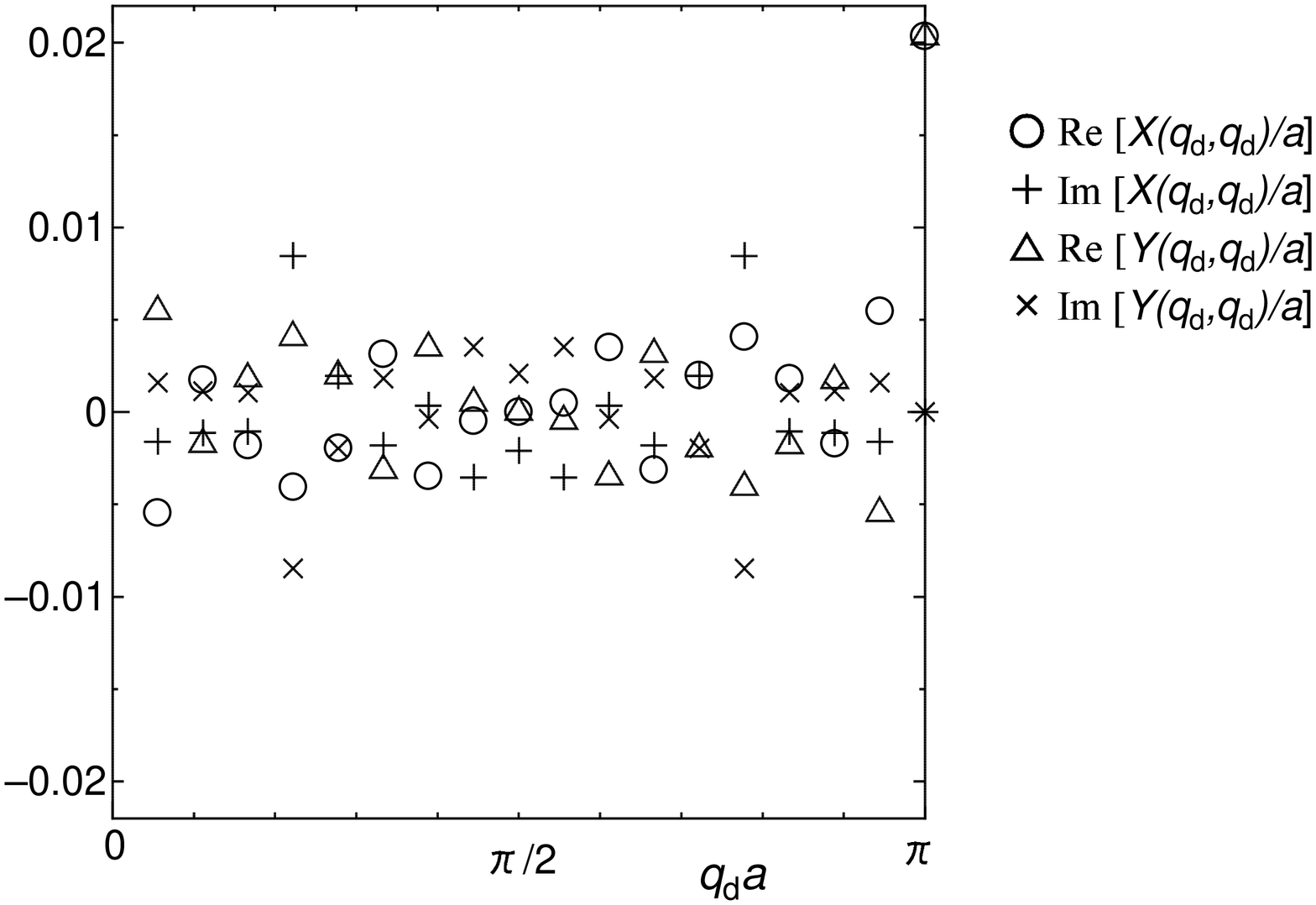,width=280pt,height=198pt}
%\centerline{\epsfbox{fq36_07.ps}}
%\vspace*{5cm}
\caption{The Fourier spectra of the distortion pattern obtained by numerical 
minimization of the total energy. As for the direction of the wave number 
$\vv{q}$, the Fourier components vanish when $q_x\ne q_y$. 
The abscissa indicates $q_{\rm d}a$ ($=q_xa=q_ya$). The system size is 
$36\times 36$. The electron-lattice coupling constant $\alpha$ is equal to 
4.0eV/\AA$^2$ which corresponds to $\lambda=0.30$.}
\label{fig2}
\end{figure}

What is characteristic in these Fourier spectra is that the amplitudes 
[$=\sqrt{|X(\vv{q}_i)|^2+|Y(\vv{q}_i)|^2}$ ($i=1,2$)] of 
two modes with $\vv{q}_1$ and $\vv{q}_2$ satisfying the condition 
$\vv{q}_1+\vv{q}_2=\vv{Q}$ are equal to each other. The meaning of this 
property will be discussed in the next section. It should be noted that 
as far as the $\vv{Q}$-components are concerned the amplitudes are 
highly isotropic in $x$ and $y$ directions. 

The electronic energy spectrum corresponding to this lattice configuration 
is different from that in the asymmetric dimerization case (Fig.~\ref{fig1}). 
There is no point on the Fermi surface where the Peierls gap vanishes. The 
behavior of the gap along the Fermi surface will be discussed later 
in detail. 

\begin{figure}[htb]
%\vspace*{-5cm}
%\epsfysize=6.0cm
\figureheight{6cm}
\hspace*{30mm}
\epsfig{file=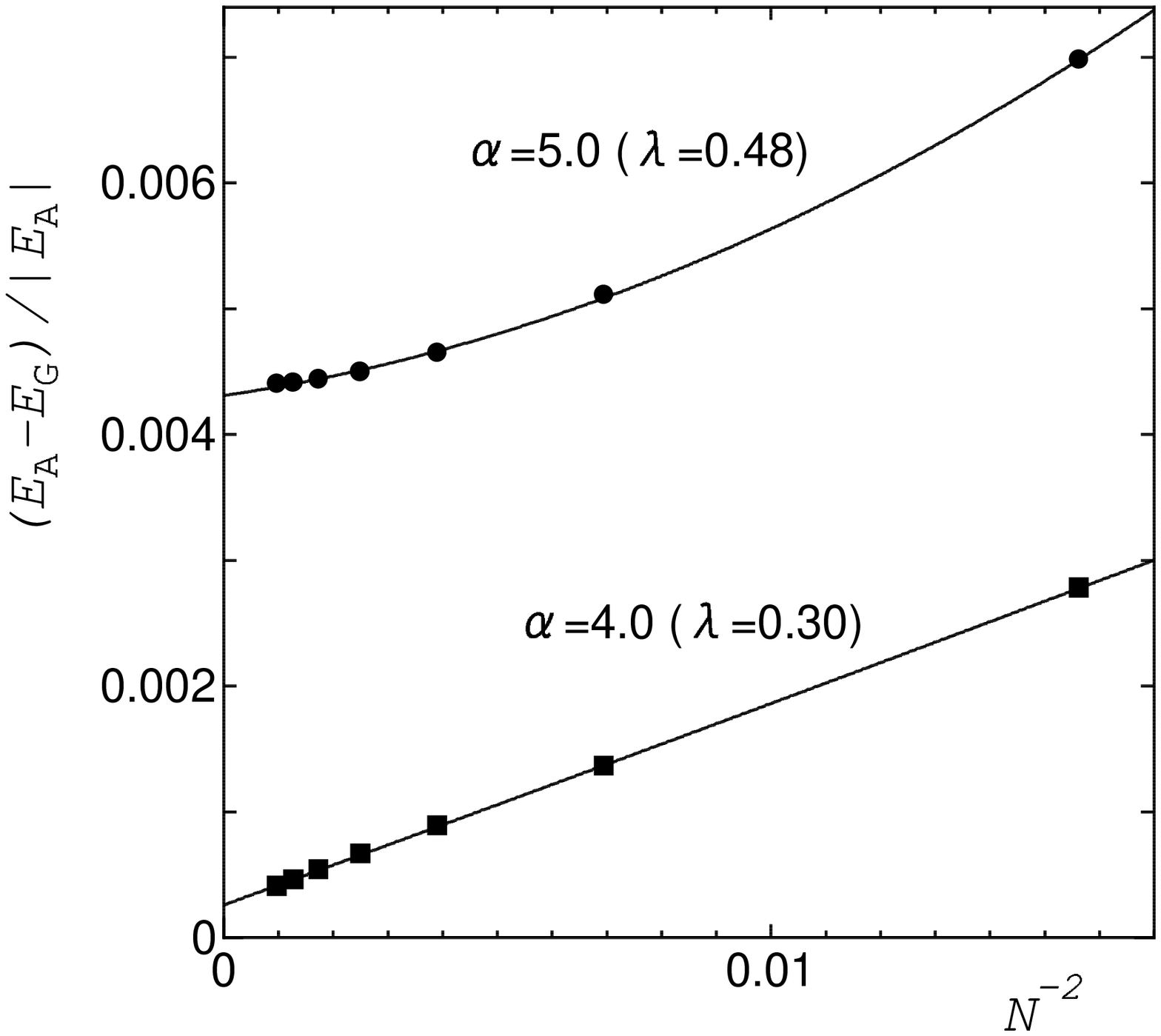,width=280pt,height=198pt}
%\centerline{\epsfbox{fq36_07.ps}}
%\vspace*{5cm}
\caption{The size-dependence of the energy difference between the asymmetrically 
dimerized state and the lowest energy state obtained by solving the self-consistent 
equations. The abscissa is the inverse of the total number of the lattice points. 
The value of $\alpha$ is indicated in the figure along with the corresponding 
value of the dimensionless coupling constant $\lambda$, other parameters being 
fixed ($t_0=2.5$eV and $K=21.0$eV/\AA$^2$). The continuous lines are fitting to 
quadratic polynomials.}
\label{fig3}
\end{figure}

In order to check whether this complicated state has a lower energy 
than the asymmetrically dimerized state as shown in Fig.~\ref{fig1}, 
we study the size dependence of the energy difference between the 
asymmetrically dimerized state (Fig.~\ref{fig1}) and the lowest energy 
state obtained by solving the self-consistent equations (\ref{eq3.1}) 
to (\ref{eq3.3}); the total energy of the former state is denoted by 
$E_{\rm A}$ and the latter by $E_{\rm G}$, both of them being negative. 
The result is summarized in Fig.~\ref{fig3}, where the energy difference 
scaled by $|E_{\rm A}|$ is plotted as a function of $N^{-2}$, the inverse 
of the total number of the lattice points, for two different values of 
$\alpha$, other parameters being fixed as mentioned before. 

From Fig.~\ref{fig3}, we may safely conclude that the energy difference remains 
finite in the thermodynamic limit, and that the lowest energy state obtained from 
the self-consistent equation has certainly a smaller energy than the asymmetrically 
dimerized state shown in Fig.~\ref{fig1}.

%%%%%%%%%%%%%%%%%%%%%%%%%%%%%%%%%%%%%%%%%%%%%%%%%%%%%%%%%%%%%%%%%%%%%%%%%%%%
%%%%%%%%%%%%%%%%%%%%%%%%%%%%%%%%%%%%%%%%%%%%%%%%%%%%%%%%%%%%%%%%%%%%%%%%%%%%

\section{Peierls Gap due to Second Order Process} %sec4

In this section we discuss why the single mode dimerization (Fig.~\ref{fig1}) 
is not the lowest energy state. The lowest energy state obtained numerically 
in the previous section involves multi-mode distortions. The Fourier analysis 
of the distortions indicates that the gap in the electronic spectrum is 
induced not only by the first order perturbation but also by the second 
order process. As discussed in \S~1, the Peierls gap in the electronic 
spectrum is formed usually by the first order process in which a state 
$|\vv{k}\rangle$ 
on the Fermi surface is coupled to another state $|\vv{k}\pm \vv{Q}\rangle$ 
on the Fermi surface by the lattice distortion with a wave vector $\vv{Q}$ 
(the nesting vector). The consideration developed in \S~2 indicates that 
the gap vanishes at special points on the Fermi surface, $(\pi/a,0)$ and 
$(0,\pi/a)$, by this first order process because of the peculiar wave 
number dependence of the electron-lattice coupling term. This fact means 
that in order to get a finite gap at those point we have to consider the 
second order process where at least two lattice distortion modes should be 
relevant. 

If two distortion modes with wave vectors $\vv{q}_1$ and $\vv{q}_2$ satisfying 
$\vv{q}_1+\vv{q}_2=\vv{Q}$ are involved in the second order process, there 
should appear matrix elements connecting one of the states $|\vv{k}\rangle$ or 
$|\vv{k}+\vv{Q}\rangle$ and one of $|\vv{k}+\vv{q}_1\rangle$ or 
$|\vv{k}+\vv{q}_2\rangle$, where the former states are on the Fermi surface. 
This indicates in turn that the two states $|\vv{k}+\vv{q}_1\rangle$ and 
$|\vv{k}+\vv{q}_2\rangle$ which are not on the Fermi surface are mixed with 
the states $|\vv{k}\rangle$ and $|\vv{k}+\vv{Q}\rangle$ on the Fermi surface 
and contribute to the formation of the gap at the Fermi surface. In such a 
situation it is natural to expect the creation of lattice distortions with 
wave vectors $\vv{q}_1-\vv{q}_2$, $\vv{q}_2-\vv{q}_1$, 
$\vv{q}_1-\vv{q}_2+\vv{Q}(=2\vv{q}_1)$ and 
$\vv{q}_2-\vv{q}_1+\vv{Q}(=2\vv{q}_2)$. In this way, many lattice distortion 
modes can be involved in the second order process. 

Based on the numerical results discussed in the previous section, we assume 
that, in the lowest energy state, only the lattice distortions with wave 
vectors parallel to $\vv{Q}$ are existing.  Namely the bond variables 
$x_{i,j}$ and $y_{i,j}$ are assumed to be expressed in the form,
\begin{eqnarray}
x_{i,j} & = & x_0(-1)^{i+j} +\sum_{0<q<\pi/a} [x_q{\rm e}^{{\rm i}qa(i+j)} 
+ {\rm c.c.}], \label{eq4.1} \\
y_{i,j} & = & y_0(-1)^{i+j} +\sum_{0<q<\pi/a} [y_q{\rm e}^{{\rm i}qa(i+j)} 
+ {\rm c.c.}]. \label{eq4.2} 
\end{eqnarray}
By this assumption the lattice degree of freedom is reduced from $N^2$ to 
$N-1$; the uniform mode with $q=0$ is excluded trivially. 
Then the original Hamiltonian can be written in the wave number representation 
as follows,
\begin{eqnarray}
H & = & \sum_{\vs{k},s} \varepsilon_{\vs{k}}c_{\vs{k},s}^\dagger c_{\vs{k},s} 
\nonumber \\
& & +\alpha\sum_{\vs{k},s}2{\rm i}(x_0\sin k_xa+y_0\sin k_ya)
c_{\vs{k}+\vs{Q},s}^\dagger c_{\vs{k},s} \nonumber \\
& & +\alpha \sum_{0<q<\pi/a} \sum_{\vs{k},s} 2\left\{ {\rm e}^{-{\rm i}qa/2} 
\left[ x_q\cos \left( k_x+{q \over 2}\right) a \right. \right. \nonumber \\
& & \left. +y_q\cos \left( k_y+{q \over 2}\right) a\right]
c_{\vs{k}+\vs{q},s}^\dagger c_{\vs{k},s}  \nonumber \\
& & +{\rm e}^{{\rm i}qa/2}\left[ x_q^*\cos 
\left( k_x-{q \over 2}\right) a \right. \nonumber \\
& & \left. \left. +y_q^*\cos 
\left( k_y-{q \over 2}\right) a \right]c_{\vs{k}-\vs{q},s}^\dagger c_{\vs{k},s}
\right\} \nonumber \\
& & +N^2{K \over 2}(x_0^2+y_0^2) \nonumber \\
& & + N^2 K\sum_{0<q<\pi/a}(|x_q|^2+|y_q|^2), 
\label{eq4.3}
\end{eqnarray}
where the vector $\vv{q}$ stands for $(q,q)$. 

Thus, in order to obtain the electronic energy spectrum in the presence of 
the lattice distortions considered above, we have to solve the eigenvalue 
problem of the following form, 
\begin{eqnarray}
\varepsilon \psi(k_x,k_y) & = & \varepsilon_{\vs{k}}\psi(k_x,k_y) \nonumber \\
& &-2{\rm i}\alpha (x_0\sin k_xa+y_0\sin k_ya)\times \nonumber \\
& & \psi(k_x-Q_x,k_y-Q_y) \nonumber \\
& & +2\alpha \sum_{0<q<\pi/a} \left\{ {\rm e}^{-{\rm i}qa/2}\left[ x_q\cos 
\left( k_x-{q \over 2}\right) a \right. \right. \nonumber \\
& & \left. +y_q\cos \left( k_y-{q \over 2}\right) a\right] \psi(k_x-q,k_y-q) 
\nonumber \\
& & + {\rm e}^{{\rm i}qa/2}\left[ x_q^*\cos 
\left( k_x+{q \over 2}\right) a \right.  \nonumber \\
& & \left. \left. +y_q^*\cos \left( k_y+{q \over 2}\right) a\right] 
\psi(k_x+q,k_y+q) \right\} , \nonumber \\
\label{eq4.4}
\end{eqnarray}
where $\psi(k_x,k_y)$ means the wave function in the wave number 
representation. It should be noted that, in the above equation, only the 
combinations of $k_x$ and $k_y$ with a constant difference are involved. 
It is not difficult to understand from this fact that the Hamiltonian 
matrix whose original size is $N^2\times N^2$ can be decomposed into $N$ 
pieces each of which has a size $N\times N$. Therefore the most general 
eigenvalue problem eq.~(\ref{eq3.1}) is decoupled into $N$ one dimensional 
problems in the present situation. Each eigenvalue equation of the form 
eq.~(\ref{eq4.4}) yields a set of $N$ eigenvalues and $N$ eigenfunctions. 
By solving $N$ sets of eigenvalue problems for one set of \{$x_q,\,y_q$\} 
and $(x_0,\,y_0)$, we end up with a set of $N^2$ eigenvalues and $N^2$ 
eigenfunctions. Sorting $N^2$ eigenvalues in increasing order, we determine 
the state index $\nu$ and attach it to each eigenvalue and eigenfunction 
as $\varepsilon_\nu$ and $\psi_\nu(k+k',k)$. If we note the periodicity 
of the wave functions in the $\vv{k}$-space, we have only to consider the 
range of $k$ and $k'$ such as $-\pi/a <k, k' \le \pi/a$. 

Next we have to discuss how to determine the set of variables, 
\{$x_q,\,y_q$\} and $(x_0,\,y_0)$. They are determined so as to minimize 
the total energy of the system as done in \S~3. The self-consistent 
equations for these variables are written in the following form,
\begin{eqnarray}
x_0 & = & -{{2{\rm i}\alpha} \over {KN^2}}{\sum_\nu}' \sum_{\vs{k}} 
\sin k_xa \psi_\nu^*(k_x+\pi/a,k_y+\pi/a)\times \nonumber \\
& & \psi_\nu(k_x,k_y) , \label{eq4.5} \\
y_0 & = & -{{2{\rm i}\alpha} \over {KN^2}}{\sum_\nu}' \sum_{\vs{k}} 
\sin k_ya \psi_\nu^*(k_x+\pi/a,k_y+\pi/a)\times \nonumber \\
& & \psi_\nu(k_x,k_y) , \label{eq4.6} \\
x_q & = & -{{2{\rm e}^{{\rm i}qa/2}\alpha} \over {KN^2}}{\sum_\nu}' 
\sum_{\vs{k}}\cos \left( k_x-{q \over 2}\right) a 
\times \nonumber \\
& & \psi_\nu^*(k_x-q,k_y-q) \psi_\nu(k_x,k_y) , \label{eq4.7} \\
y_q & = & -{{2{\rm e}^{{\rm i}qa/2}\alpha} \over {KN^2}}{\sum_\nu}' 
\sum_{\vs{k}}\cos \left( k_y-{q \over 2}\right) a 
\times  \nonumber \\
& & \psi_\nu^*(k_x-q,k_y-q) \psi_\nu(k_x,k_y) , \label{eq4.8}
\end{eqnarray}
where the sum over $\nu$ is restricted to the occupied states. We have 
omitted complex conjugate forms of eqs.~(\ref{eq4.7}) and (\ref{eq4.8}).
It should be noted that for each $\nu$ the corresponding wave function 
is finite only when $k_x-k_y=$const., though this constant depends on 
$\nu$. 

In the present calculation scheme, we have only to diagonalize $N$ 
different $N\times N$ matrices instead of a single $N^2\times N^2$. 
%Namely the two dimensional problem is reduced into multiple one 
%dimensional problems. 
This makes the computational load much lighter, and therefore we 
can treat far larger system than in the case where we deal with two 
dimensional problems directly. 

\begin{figure}[htb]
%\vspace*{-5cm}
%\epsfysize=6.0cm
\figureheight{6cm}
\hspace*{30mm}
\epsfig{file=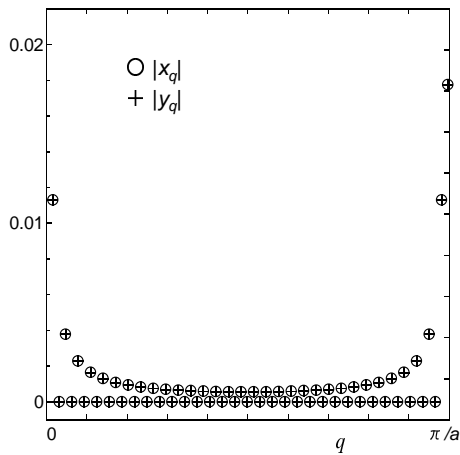,width=280pt,height=198pt}
%\centerline{\epsfbox{fq36_07.ps}}
%\vspace*{5cm}
\caption{The $q$ dependences of the amplitudes $|x_q|$ and $|y_q|$. The 
values of $|x_0|$ and $|y_0|$ are also plotted. The system size is $N=128$, 
and the coupling constant is $\alpha=4.0$eV/\AA \ ($\lambda =0.30$). }
\label{fig4}
\end{figure}

\begin{figure}[htb]
%\vspace*{-5cm}
%\epsfysize=6.0cm
\figureheight{6cm}
\hspace*{30mm}
\epsfig{file=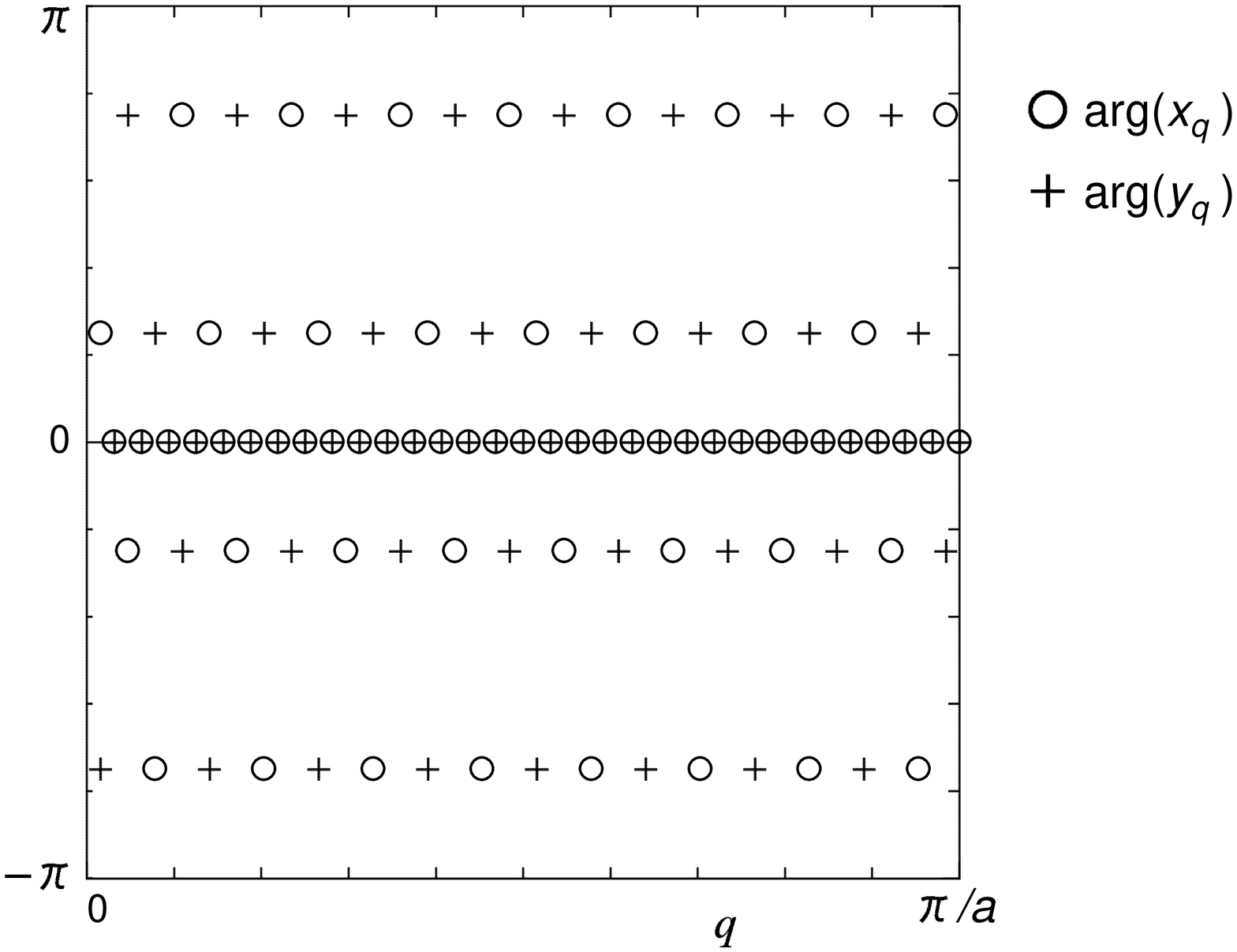,width=280pt,height=198pt}
%\centerline{\epsfbox{fq36_07.ps}}
%\vspace*{5cm}
\caption{The $q$ dependences of the phases arg$(x_q)$ and arg$(y_q)$. 
The values of arg$(x_0)$ and arg$(y_0)$, which are zero in this example, 
are also given for comparison. The system size is $N=128$, and the coupling 
constant is $\alpha=4.0$eV/\AA \ ($\lambda =0.30$). }
\label{fig5}
\end{figure}

In Figs.~\ref{fig4} and \ref{fig5} we show the wave number dependences of 
the amplitudes and the phases of $x_q$'s and $y_q$'s which have been obtained 
by solving the self-consistent equations for the system size $N=128$ (the 
total number of lattice points $=128\times 128$) and the coupling constant 
$\alpha = 4.0$ eV/\AA \ ($\lambda=0.30$). The initial condition used in this 
iterative calculation is Re\ $x_q={\rm Im}\ x_q=x_{\rm in}$ (some constant of 
the order of $10^{-2}a$) for $q=2\pi/Na$, $x_q=0$ for other $q$ and 
$y_q=0$ for all $q$, where $q$ is restricted in the region $0<q<\pi/a$, 
and $x_0=y_0=x_{\rm in}$. The final results does not 
depend on the values of $x_{\rm in}$.  The results 
for different types of initial conditions will be discussed later.

The amplitudes of $x_q$ and $y_q$ are completely equal to each other and are 
found to vanish at $q$ values which are even integer times $2\pi/Na$, though, 
as a matter of course, $q=\pi/a$ is exceptional. As for the phases, we find 
the relation arg($x_{q_1}$)+arg($x_{q_2}$)=arg($y_{q_1}$)+arg($y_{q_2}$)=0 
mod($\pi$) for $q_1$ and $q_2$ satisfying $q_1+q_2=\pi/a$. This will be 
reasonable if we remind us the second order perturbation mechanism of the 
Peierls gap formation. 

\begin{figure}[htb]
%\vspace*{-5cm}
%\epsfysize=6.0cm
\figureheight{6cm}
\hspace*{30mm}
\epsfig{file=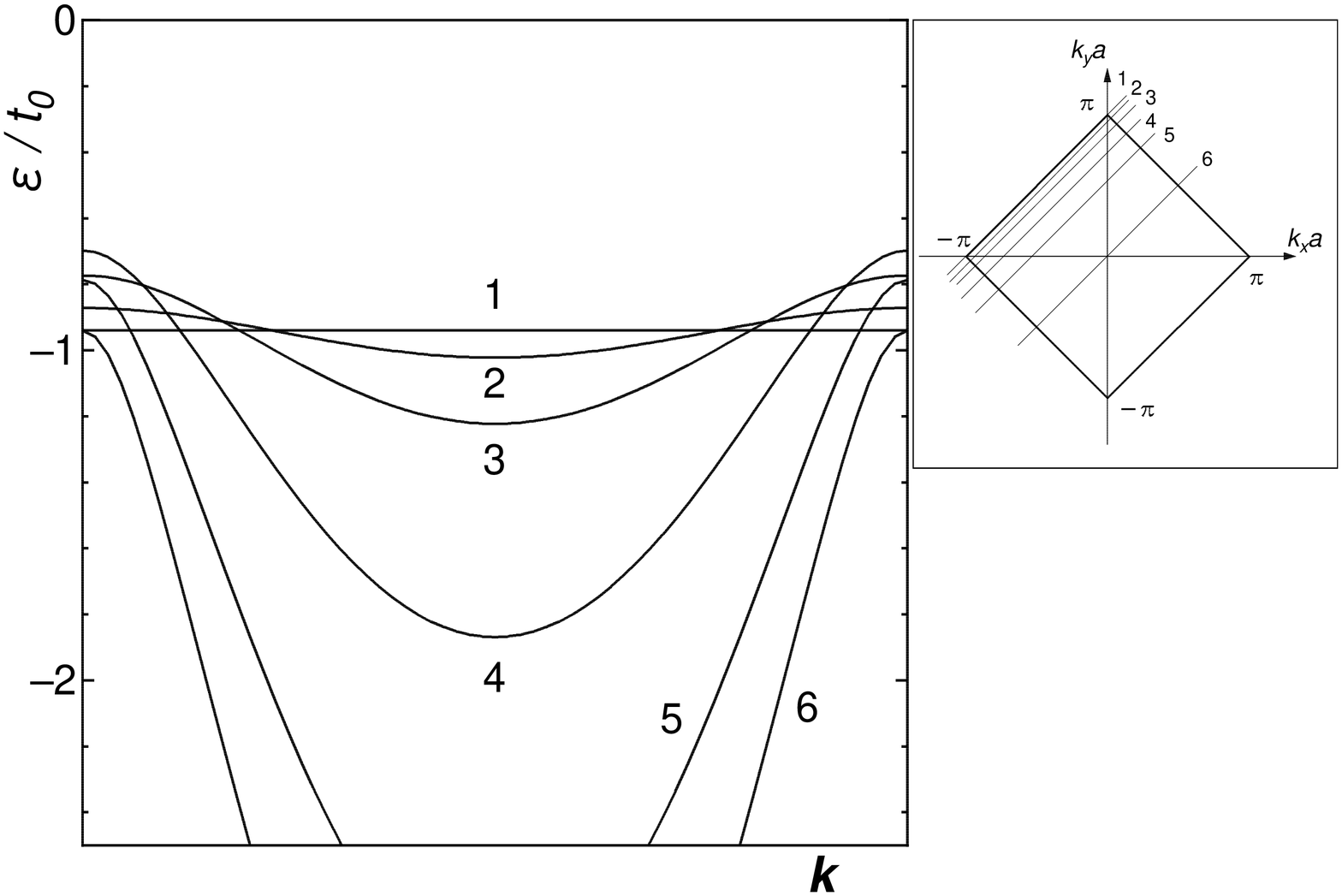,width=280pt,height=198pt}
%\centerline{\epsfbox{fq36_07.ps}}
%\vspace*{5cm}
\caption{The dispersion relations of occupied levels for $\alpha =6.0$eV/\AA 
\ ($\lambda=0.69$) and the system size $128\times 128$. The wave numbers are 
changed along the lines indicated in the sub-figure. }
\label{fig6}
\end{figure}

Once we know the values of \{$x_q$\}, \{$y_q$\}, $x_0$ and $y_0$, we can 
obtain $N$ eigenenergies for each set of $N$ electronic wave vectors. 
Assuming that the energy of the lower (higher) band is an increasing 
(decreasing) function of the distance from the line $k_x+k_y=0$, we can 
assign the energy versus $\vv{k}$ relation. The dispersion relation obtained 
in this way is shown in Fig.~\ref{fig6} for the case with $\alpha=6.0$eV/\AA 
\ ($\lambda=0.69$).~\cite{foot2} Only the lower band which is fully occupied in 
the ground state is shown along the lines depicted in the sub-figure. Because 
of the electron-hole symmetry of the system the dispersions of the unoccupied 
levels are the same as those shown in Fig.~\ref{fig6} except for the sign of 
energy. 
 
Thus by assigning 
the dispersion relation for all the points in the $\vv{k}$-space, we can now 
see the gap structure on the Fermi surface. As will be clear from 
Fig.~\ref{fig6}, the gap is constant along the line $(-\pi/a,0)-(0,\pi/a)$. 
On the other hand the gap structure along the line $(0,\pi/a)-(\pi/a,0)$ 
is not so simple as will be found from Fig.~7, where the dispersion curves 
of upper and lower bands along the line $(0,\pi/a)-(\pi/a,0)$ are shown 
for the cases with a couple of different values of the electron-lattice 
coupling constant. 

\begin{figure}[htb]
%\vspace*{-5cm}
%\epsfysize=6.0cm
\figureheight{6cm}
\hspace*{30mm}
\epsfig{file=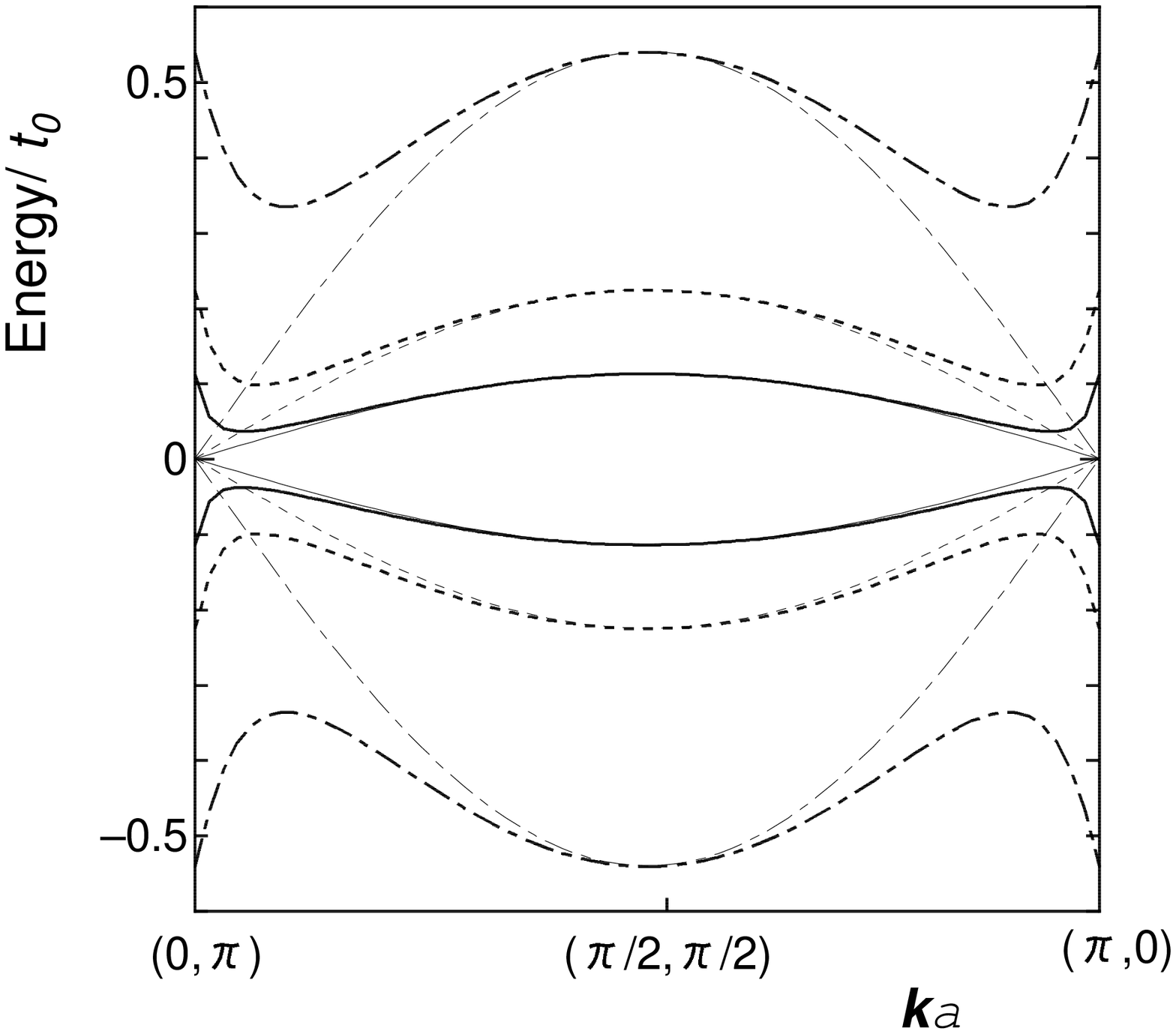,width=280pt,height=198pt}
%\centerline{\epsfbox{fq36_07.ps}}
%\vspace*{5cm}
\caption{The dispersions of the highest occupied and lowest unoccupied levels 
along the line $(0,\pi/a)-(\pi/a,0)$. Three different values of 
electron-lattice constant are considered, $\alpha [{\rm eV/\AA}]=4.0$ 
($\lambda= 0.30$); continuous thick line, 4.4 (0.37); dotted line and 
5.2 (0.52); dash-dotted line. The similar dispersions for the asymmetric 
dimerized state (Fig.~1) are also shown by the same type thinner lines. }
\label{fig7}
\end{figure}

In contrast to the asymmetrically dimerized case (Fig~\ref{fig1}), the gap 
does not vanish at any point, taking almost the same value as that at 
$(\pi/2a,\pi/2a)$. It will be noteworthy that the gap is not minimum at 
$\vv{k}=(0,\pi/a)$ or $(\pi/a,0)$ and that the multi-mode effect is almost 
negligible around the point $\vv{k}=(\pi/2a,\pi/2a)$. The position of the 
gap minimum depends on the coupling constant. In the same figure, the 
dispersions in the asymmetrically dimerezed case for each coupling constant 
are shown for comparison. 

\begin{figure}[htb]
%\vspace*{-5cm}
%\epsfysize=6.0cm
\figureheight{6cm}
\hspace*{30mm}
\epsfig{file=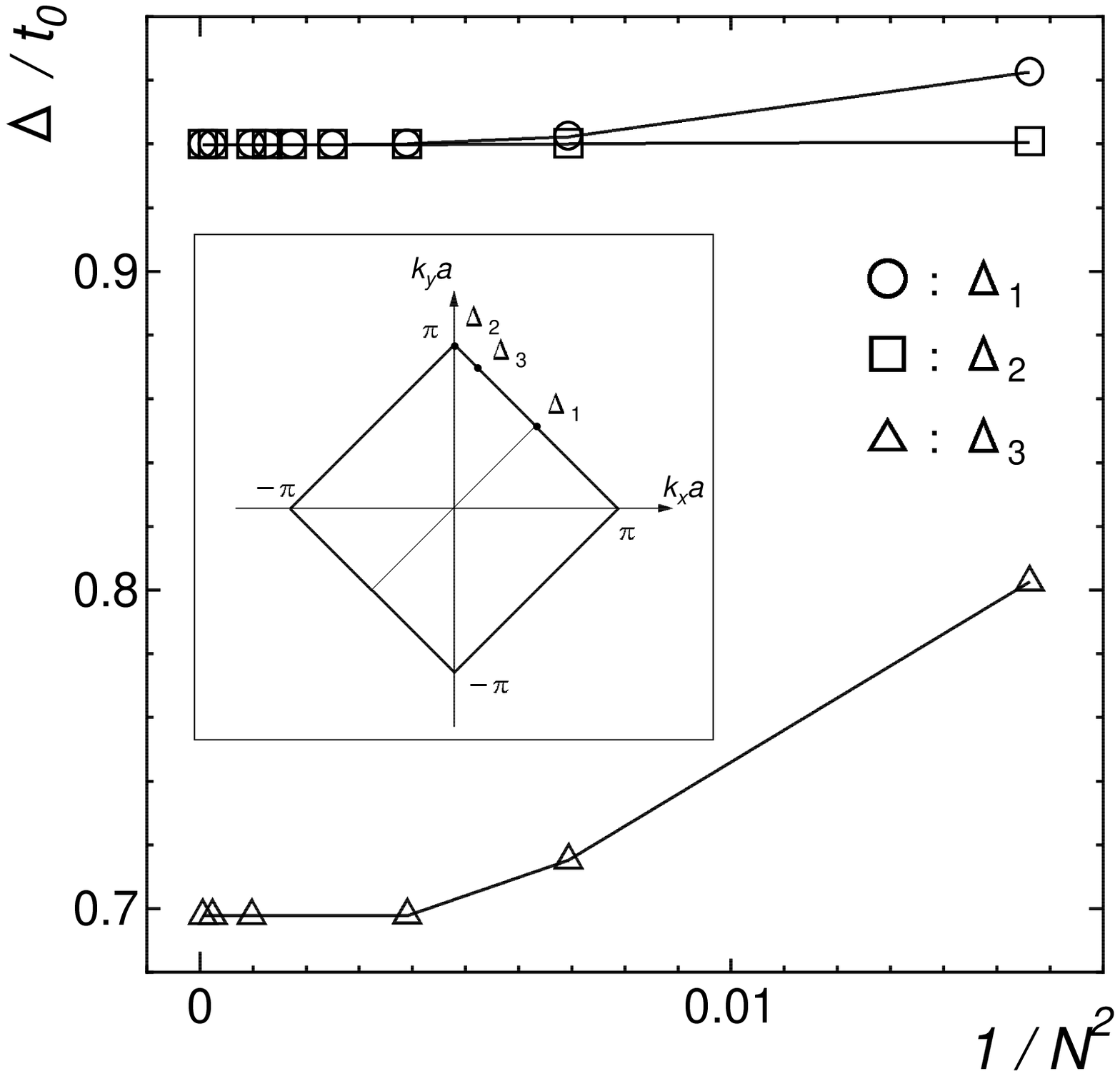,width=280pt,height=198pt}
%\centerline{\epsfbox{fq36_07.ps}}
%\vspace*{5cm}
\caption{The system size dependence of the gap $\Delta$ in the electronic 
spectrum at different points on the line $(0,\pi/a)-(\pi/a,0)$. The coupling 
constant is fixed at $\lambda=0.69$.}
\label{fig8}
\end{figure}

In order to see the size dependence of the gap $\Delta$, we plot in 
Fig.~\ref{fig8} the gap at three different points on the line 
$(0,\pi/a)-(\pi/a,0)$, as indicated in the inset, for a coupling constant 
$\lambda = 0.69$ ($\alpha =6.0$eV/\AA). From this figure we find that 
$N=128$ belongs to the large size limit as far as the behavior of the 
gap concerns and that in the large size limit the gaps $\Delta_1$ and 
$\Delta_2$ at $(\pi/2a,\pi/2a)$ and $(0,\pi/a)$ are indistinguishable 
though the value of $\Delta_3$ remains smaller than $\Delta_1$ and 
$\Delta_2$. For smaller values of the coupling constant, a similar tendency 
is found, but the large size limit is seen only at larger system sizes, 
although $N=128$ is sufficiently large even for $\alpha = 4.0$eV/\AA. 

As indicated by the results shown above, the lowest energy state obtained 
here is not completely symmetric with respect to $x$ and $y$ directions. 
This fact means that there exist at least two degenerate 
states where the roles of $x$ ant $y$ are interchanged. 

In order to check other degeneracy of the lowest energy states, we have 
studied what kind of distortion patterns could be obtained when we change 
the initial distortions in the iterative calculation. As a result, we 
got many different distortion patterns with the same energy as that of 
the distortion pattern shown in Figs.~\ref{fig4} and \ref{fig5}. Some 
are different only in the phases of the Fourier components of the distortion. 
Some show completely different behaviors. Among them there is a pattern where 
the nonvanishing Fourier components are $x_{\pi/2a}$, $y_{\pi/2a}$, $x_0$ 
and $y_0$ and other components are zero. In these degenerate states, not only 
the total energy but also the electronic part and the lattice potential part 
are also the same, respectively. At the moment we cannot say how many states 
are degenerate. Detailed analysis of the degeneracy problem is left for a 
future work.

%%%%%%%%%%%%%%%%%%%%%%%%%%%%%%%%%%%%%%%%%%%%%%%%%%%%%%%%%%%%%%%%%%%%%%%%%%%%
%%%%%%%%%%%%%%%%%%%%%%%%%%%%%%%%%%%%%%%%%%%%%%%%%%%%%%%%%%%%%%%%%%%%%%%%%%%%
\section{Summary and Discussion} %sec5

The lowest energy state of a two dimensional electron-lattice system 
with a half-filled electronic band and with a square lattice structure 
is studied within the SSH-type model 
extended to two dimensions. On the contrary to the previous common 
understanding that only the lattice distortion with the nesting 
vector $\vv{Q}=(\pi/a,\pi/a)$ is frozen in the lowest energy 
state,\cite{hirsch1} many modes are found to be frozen in the real 
lowest energy state. The state discussed by Tang and Hirsch\cite{hirsch1}
might be a local minimum but it is not the absolute minimum of energy. 
This is because the Peierls gap vanishes at 
$(\pi/a,0)$ and $(0,\pi/a)$ due to the wave number dependence of the 
electron-lattice coupling term. We have pointed out that the second 
order perturbation mechanism of the Peierls gap formation is important. 
Numerical minimization of the total energy leads to the conclusion that 
many modes having the wave number parallel to $\vv{Q}$ 
contribute to the formation of the gap. Assuming this is the case, we can 
reduce the two-dimensional problem into one-dimensional problems by 
using the wave number representation of the Hamiltonian as discussed in \S~4. 
In this formulation it is possible to treat far larger system sizes than in 
the case where we search a minimum energy state of the two-dimensional 
system directly. The electronic energy dispersion and the gap structure 
on the Fermi surface have been analyzed. Although the wave number dependences 
of the amplitudes and phases of condensed modes look to show a certain 
symmetry in the $x$ and $y$ directions, the electronic structures are 
not necessarily symmetric. 

What we have shown in this paper is not the unique lowest energy state. 
A preliminary study indicates there might be infinite number of degenerate 
ground states. The degree of degeneracy is not known at the moment. It is 
not clear also what kind of symmetry is relevant to this degeneracy. 
Detailed study of the degeneracy problem is left for the future work. 
Nevertheless it will be worthwhile to mention that this type of degeneracy 
of the ground state yields the possibility of the formation of domain 
walls like the solitons in the one-dimensional systems where the number of 
degenerate ground state is only two.\cite{SSH,rice,TLM} This kind of 
domain walls connecting two different ground states may supply interesting 
properties of the two-dimensional electron-lattice systems just as 
the charged and neutral solitons in polyacetylene did.

%%%%%%%%%%%%%%%%%%%%%%%%%%%%%%%%%%%%%%%%%%%%%%%%%%%%%%%%%%%%%%%%%%%%
%%%%%%%%%%%%%%%%%%%%%%%%%%%%%%%%%%%%%%%%%%%%%%%%%%%%%%%%%%%%%%%%%%%%
\section*{Acknowledgments}

The authors are grateful to Professor Y. Wada for useful comment on 
existing references. 

%%%%%%%%%%%%%%%%%%%%%%%%%%%%%%%%%%%%%%%%%%%%%%%%%%%%%%%%%%%%%%%%%%%%%
%%%%%%%%%%%%%%%%%%%%%%%%%%%%%%%%%%%%%%%%%%%%%%%%%%%%%%%%%%%%%%%%%%%%%
\def\jpsj{J. Phys. Soc. Jpn. }
\def\ptp{Prog. Theor. Phys. }
\def\prl{Phys. Rev. Lett. }
\def\pr{Phys. Rev. }

%\end{multicols}
%\end{large}


\begin{thebibliography}{99}
%%%%%%%%%%%%%%%%%%%%%%%%%%%%%
\bibitem{peierls}
R.E. Peierls: {\it Quantum theory of Solids} (Clarendon Press, Oxford, 1955). 
\bibitem{machida87}
K. Machida and M. Kato: \pr B{\bf 36} (1987) 854.
\bibitem{hirsch1}
S. Tang and J.E. Hirsch: \pr B{\bf 37} (1988) 9546.
\bibitem{scal89}
R.T. Scalettar, N.E. Bickers, and D.J. Scalapino: \pr B{\bf 3?} (1989) ???
\bibitem{mazumdar}
S. Mazumdar: \pr B{\bf 39} (1989) 12324. 
%%%%%%%%%%%%%%%%%%%%%%%%%%%%%
\bibitem{SSH}
W. P. Su, J. R. Schrieffer, and A. J. Heeger: 
\prl {\bf 42} (1979) 171, \pr {\bf B22} (1980) 2099.
\bibitem{stafstrom}
S. Stafstr\"om and K.A. Chao: \pr B{\bf 29} (1984) 7010; 
{\it ibid} {\bf 30} (1984) 2098.
\bibitem{TO86}
A. Terai and Y. Ono: \jpsj {\bf 55} (1986) 213.
%%%%%%%%%%%%%%%%%%%%%%%%%%%%%
\bibitem{footnote1}
Because of square lattice symmetry, the wave vectors for which the Fourier 
components are finite can be $(q_{\rm d},-q_{\rm d})$ or 
$(-q_{\rm d},q_{\rm d})$.  We do not distinguish these cases from the case 
with $(q_{\rm d},q_{\rm d})$. 

\bibitem{foot2}
We have taken a larger coupling constant for showing the dispersion, since 
the detailed structures can be seen more clearly than those for smaller 
values of the coupling constant. 

%%%%%%%%%%%%%%%%%%%%%%%%%%%%%
\bibitem{rice}
M.J. Rice: Phys. Lett. {\bf 71A} (1979) 152.
\bibitem{TLM}
H. Takayama, Y.R. Lin-Liu, and K. Maki: \pr  B{\bf 21} (1980) 2388.

\end{thebibliography}
\end{document}